\documentclass[
    aps,prb,10pt,
    twocolumn,reprint,
    superscriptaddress,floatfix,amsmath,amssymb
]{revtex4-2}

\usepackage{graphicx}% Include figure files
\graphicspath{{}}
\usepackage[caption=false]{subfig}

\usepackage{dcolumn}% Align table columns on decimal point
\usepackage{bm}% bold math
\usepackage{hyperref}% add hypertext capabilities

\usepackage{color}
\usepackage{siunitx}
\usepackage{algorithm}
\usepackage{algpseudocode}
\usepackage{braket}
\usepackage[export]{adjustbox}
\usepackage{placeins}
\usepackage{xcolor}
\newcommand{\ignore}[1]{}

\begin{document}

\title{Heating and cooling in self-consistent many-body simulations}

\def\umphys{%
    Department of Physics, University of Michigan,
    Ann Arbor, MI 48109, USA
}%

\author{Yang Yu}
\affiliation{\umphys}

\author{Sergei Iskakov}
\affiliation{\umphys}

\author{Emanuel Gull}
\affiliation{\umphys}

\date{\today}

\begin{abstract}
%Motivation
%Problem
%Method 
%Results
%Conclusions

We present a temperature extrapolation technique for self-consistent many-body methods, which provides a causal starting point for converging to a solution at a target temperature. The technique employs the Carathéodory formalism for interpolating causal matrix-valued functions and is applicable to various many-body methods, including dynamical mean field theory, its cluster extensions, and self-consistent perturbative methods such as the self-consistent GW approximation. We show results that demonstrate that this technique can efficiently simulate heating and cooling hysteresis at a first-order phase transition, as well as accelerate convergence.
\end{abstract}

\maketitle

\section{Introduction}
When conducting experiments in condensed matter physics, it is common practice to investigate the temperature dependence of observables while keeping other parameters constant. A particular example are  specific heat or transport measurements, which are often used as a preliminary probe to identify intriguing temperature-dependent behavior. For example, in a first-order  coexistence regime, heating and cooling curves may reveal history-dependent hysteresis.

In theoretical calculations using self-consistent finite-temperature field theories, %one can perform a calculation from a converged neighboring point to simulate a gradual change of a control parameter and study history-dependent behavior if parameters other than temperature are changed. However, 
changing the temperature of a system is generally not practical as it causes a shift of Matsubara frequencies \cite{abrikosov1975a,mahan2000}. Extrapolating the lowest Matsubara frequency during cooling can be problematic, resulting in non-causal solutions. Therefore, implementing heating and cooling protocols in self-consistent finite-temperature field theories, in analogy to heating and cooling measurements in experiment,  is an unresolved issue.

Self-consistent  finite-temperature simulation methods include non-perturbative techniques, such as the dynamical mean field theory (DMFT) \cite{goedecker1996} and its cluster variants \cite{hettler1998, lichtenstein2000a,kotliar2001a,maier2005a};  self-consistent perturbative methods, such as  GW \cite{hedin1965,aryasetiawan1998,kutepov2009,kutepov2020,yeh2022a,yeh2022b}, second-order perturbation theory \cite{dahlen2005, phillips2014,rusakov2016,iskakov2019a}, fluctuation-exchange or T-matrix approximations \cite{bickers1989a,iskakov2022b}, as well as bold-line diagrammatic Monte Carlo methods \cite{prokofev2007,gull2010a,gull2011c,cohen2014}; and combinations of embedding theories with perturbation theory \cite{biermann2003,kotliar2006a,zgid2017}.

In these methods, solutions are obtained through an iterative process that involves starting with an initial guess and continuing the process until self-consistency is achieved. The number of iterations required to reach convergence is directly related to how close the starting point is to the iteration fixed point. A `good' starting point can significantly reduce the computational effort required for the simulation, whereas a `bad' starting point may result in iterations that diverge, iterate in limit cycles, or even converge to unphysical fixed points~\cite{kozik2015a}. In parameter regimes where first-order coexistence occurs, multiple physical fixed points may exist \cite{georges1996a,kotliar2004,macridin2006,gull2009,walsh2019}.

This paper addresses the challenge of generating better starting points and implementing heating and cooling protocols in self-consistent many-body methods. It presents a solution that guarantees a causal starting point by using a converged solution at a different temperature. The proposed method relies on the Carath\'{e}odory formalism \cite{fei2021a} for interpolating causal matrix-valued functions. Originally developed for analytic continuation of matrix-valued Matsubara functions to real frequencies, this formalism can be extended to temperature extrapolation, which involves evaluating an interpolant at different Matsubara frequencies. We demonstrate the effectiveness of this approach in obtaining improved starting points in the context of DMFT and real-materials perturbation theory. Additionally, we examine heating and cooling hysteresis in the context of a first-order phase transition.

The paper proceeds as follows. In Sec.~\ref{sec:method} we introduce the formalism. Sec.~\ref{sec:impl} contains a brief description of a pedagogical implementation of the method which we provide as a supplement. Sec.~\ref{sec:results} contains results for temperature extrapolation, convergence acceleration, and first-order hysteresis. Finally, Sec.~\ref{sec:conclusions} contains our conclusions.

% \begin{figure*}[tbh]
%     \centering
%     \includegraphics[width=1\linewidth]{sketch.pdf}
%     \caption{Schematic representation of the interpolation procedure for a matrix-valued finite-temperature Matsubara function $G(\mathrm{i}\omega_n)$. See text for detailed description.}
%     \label{fig:sketch}
% \end{figure*}

\section{Method}\label{sec:method}
The central object of this paper is a causal matrix-valued fermionic Matsubara function which may represent a Green's function, a self-energy, or a cumulant~\cite{stanescu2006a}. The Matsubara function is expressed as a three-dimensional tensor, $G_{ij}(\mathrm{i}\omega_n)$, which associates a matrix (identified by the indices $i$ and $j$) with every Matsubara frequency $\omega_n = (2n+1)\pi/\beta$ ($n$ denotes an integer, $\beta$ the inverse temperature).  Its continuation to the complex plane \cite{fei2021a} is denoted as $G_{ij}(\tilde{z})$. %The corresponding spectral function $A(\omega)$, with $\omega$ a real frequency, is defined as $A(\omega)=-\frac{1}{\pi}\lim_{\eta \to 0^{+}}\operatorname{Im} G_{ij}(\tilde{z} = \omega + \mathrm{i} \eta)$. $A(\omega)$ is a positive semidefinite matrix for any frequency $\omega$, due to the fact that $\mathrm{i} G(\tilde{z})$ is a Carath\'{e}odory function~\cite{Fei21A}.%, a matrix-valued extension of Nevanlinna functions~\cite{Fei21B}.
%\ignore{$G$ is a Carath\'{e}odory function, {\it i.e.}, the imaginary part of $-G(\tilde z)$ is a postive semi-definite matrix for any $\tilde z$ in the upper half of the complex plane.}
$\mathrm{i} G(\tilde{z})$ is a Carath\'{e}odory function (up to a convention-dependent minus sign), such that  $[\mathrm{i}G(\tilde{z})+(\mathrm{i} G(\tilde z))^{\dagger}]/2$ is a positive semi-definite matrix for any $\tilde z$ in the upper half of the complex plane.

The matrix structure of $G$ depends on the application. Matrices may be scalar, diagonal, block-diagonal, or fully dense. The scalar case typically appears in single-site single-orbital cases, such as single-site DMFT. The diagonal case is often encountered in momentum-space simulations, such as cluster DMFT~\cite{hettler1998,lichtenstein2000a,kotliar2001a,maier2005a}. The general multi-orbital case with dense matrices typically appears in real-materials ab-initio calculations~\cite{kotliar2006a,zgid2017}.

The  frequency-dependence of $G$ is known at discrete Matsubara frequencies $\omega_n$. In the simplest case, the frequencies are uniformly spaced positive fermionic Matsubara frequencies $\omega_n = (2n+1)\pi/\beta$,  $n=0,1, \cdots N-1$. More generally, data is provided on a non-uniform frequency grid using Chebyshev \cite{gull2018}, intermediate representation (IR) \cite{shinaoka2017}, Legendre \cite{boehnke2011}, spline \cite{kananenka2016}, or discrete Lehmann representation \cite{kaye2022} schemes with associated Fourier transforms \cite{li2020a,kaltak2020}. For both uniform and non-uniform grids, the derivations in this section remain the same, with $n$ denoting the index that enumerates the Matsubara frequencies $\omega_{n}$.
Bosonic Matsubara response functions $\Pi(\tilde z)$ are not directly related to Carath\'{e}odory functions. However, $\Pi(\tilde z)\tilde z$ and  $\Pi(\tilde z)/\tilde z^{*}$ correspond to Carath\'{e}odory functions~\cite{fei2021e}, and~\textcite{nogaki2023} recently investigated a related mapping of bosonic functions. %\textcolor{magenta}{I'm not sure if we can say their mapping is related to the mapping of $\Pi(\tilde z)\tilde z$ or  $\Pi(\tilde z)/\tilde z^{*}$ because their mapping is implicit which may not be directly written in terms of $z$ without integration (see Verification section). I tried to understand their mapping from the viewpoint of the series expansion in terms of $z$ but failed to do that. The direct relation between the two ways of mapping is still ambiguous to me. If the context here only means they both constructed an auxiliary function that is related to Nevanlinna or Carathéodory functions, I agree with the context here but we may need to reformulate the sentence?}

\ignore{Extrapolating in temperature requires the transformation of one Matsubara grid to another one corresponding to a different temperature, while maintaining the analytic structure and causality of $G$. This requires evaluating the function at the Matsubara frequency points corresponding to the new temperature. We propose to construct a causal matrix-valued interpolant through $G(\mathrm{i} \omega_n)$ and evaluating it at the Matsubara points corresponding to the changed temperature grid.} 
Extrapolating in temperature requires the transformation of one Matsubara grid to another one corresponding to a different temperature. We propose to construct a causal matrix-valued interpolant through $G(\mathrm{i} \omega_n)$ and evaluating it at the Matsubara points corresponding to the changed temperature grid to realize the temperature extrapolation. Our approach assumes that the causal matrix-valued functions $G(\tilde{z})$ at two proximate temperatures, $\beta$ and $\beta'$, share a similar analytical structure. This allows us to approximate the unknown causal function at $\beta'$ using known causal function at $\beta$. Subsequently, the unknown Matsubara function at $\beta'$ can then be approximated by evaluating the interpolant of the causal function at $\beta$ on the Matsubara grid corresponding to $\beta'$. The approximated Matsubara function can later serve as the starting point for the self-consistent calculation at $\beta'$. From the expression
\begin{equation}
    G(\tilde{z})=\int \mathrm{d} \omega \frac{A(\omega)}{\tilde{z}-\omega},
\end{equation}
we observe that the temperature dependence of $G(\tilde{z})$ arises solely from the spectral function $A(\omega)$. Although extrapolating $A(\omega)$ along the temperature axis is possible, it is not implemented in our current approach. We find that for the Matsubara functions, the variation in Matsubara grids for different temperatures usually has a more significant impact on its value compared to the changes in the spectral function $A(\omega)$. As a result, our temperature extrapolation approach works effectively at least for nearby temperatures. In fact, empirically, our extrapolation approach performs well even for relative large temperature differences, as long as the system remains in the same phase, as demonstrated in  Sec.~\ref{subsec:comparison}. The method's algorithmic steps follow those presented for analytic continuation in Ref.~\cite{fei2021a}, with the only distinction that the interpolant is evaluated on the imaginary axis, rather than just above the real axis. The major steps of the algorithm are described below.

%For a Carath\'{e}odory function $\tilde{F}(\tilde{z})=\mathrm{i} G(\tilde{z})$, where $\operatorname{Im}\tilde{z}>0$
For a Carath\'{e}odory function $\tilde{F}(\tilde{z})=\mathrm{i} G(\tilde{z})$, $\tilde z$ in the upper half of the complex plane, we assume the values of $\tilde{F}$ are known at a set of points $\{\tilde{z}_{n}|  n=0,1, \cdots, N-1\}$. A M\"{o}bius transform
\begin{equation}
z = \frac{\tilde{z}-\mathrm{i}}{\tilde{z}+\mathrm{i}}
\label{eqn:Mobius}
\end{equation}
% accompanied by its inverse $\tilde{z}=(z+\mathrm{i})/(z-\mathrm{i})$,
maps the upper half of the complex plane to the unit disk, the real axis to the unit circle, and the Matsubara points to points onto the real axis. Since a Carath\'{e}odory function $F$ is a function defined on an open subset of the complex plane and exhibits the property $(F+F^{\dagger})/2$ being positive semi-definite, $C(z) = \tilde{F}(\tilde{z})$ then defines a Carath\'{e}odory function on the unit disk $|z|<1$. The function $C(z)$ is known at {a set of points $\{z_{n}|  n=0,1, \cdots, N-1\}$ with the values of
\begin{align}
C_{n}\equiv C(z_{n}) = \tilde{F}(\tilde{z}_{n}).
\label{eqn:C_F}
\end{align}
\ignore{with $n=0,1, \cdots, N-1$.}

The Cayley transform \cite{cayley1846} and its inverse, defined as
\begin{align}
    S(z) &= \left[I-C\left(z\right)\right] \left[I+C\left(z\right)\right]^{-1},
    \label{eqn:S_C}\\
    C(z) &= \left[I+S\left(z\right)\right]^{-1} \left[I-S\left(z\right)\right],
    \label{eqn:C_S}
\end{align}
map the Carath\'{e}odory function $C(z)$ to a matrix-valued Schur function $S(z)$~\cite{schur1918} and vice versa. A Schur function is defined as a function $S(z)$ on the unit disk with the property $||S(z)||\leq 1$, where the matrix norm $|| \cdot ||$ is defined as the largest singular value of $S$, or equivalently the largest eigenvalue of $(S S ^{\dagger})^{1/2}$. 

The problem of interpolating the Carathéodory function $C(z)$ is thus converted into the problem of determining an interpolant for the Schur function $S(z)$, where $S(z)$ is known at $N$ discrete points, given as $S_{n} \equiv S(z_n)$, with $n=0,1, \cdots, N-1$.

To identify the interpolant of the Schur function $S(z)$, we proceed to find a set of Schur functions $S^{0}(z), S^{1}(z), \cdots, S^{m}(z), \cdots, S^{N}(z)$, where each Schur function $S^{m}(z)$ is known at $N-m$ points $z=z_{m},z_{m+1}, \cdots, z_{N-1}$ with the values of $S^{m}_{n} \equiv S^{m}(z_{n})$. $S^{N}(z)$ is an arbitrary Schur function without any constraints. We require $S^{0}_{n} = S_{n}$ such that $S^{0}(z)$ serves as the desired interpolant for the Schur function $S(z)$.

To establish a connection between the two Schur functions $S^{m}(z)$ and $S^{m+1}(z)$, a function $B^{m}(z)$ is introduced, enabling a progressive determination of $S^{0}(z)$ starting from $S^{N}(z)$~\cite{fei2021a}. The relationship between $B^{m}(z)$ and $S^{m}(z)$ is given by:
\begin{equation}
    \begin{aligned}
        \frac{|z_{m}|(z_{m}-z)}{z_{m}(1-z_{m}^{*}z)}  B^{m} (z) 
        & =  L^{m} [S^{m}(z)-S^{m}_{m} ] \\
        & \times [I - {S_{m}^{m}}^{\dagger} S^{m}(z)]^{-1} R^{m}
    \end{aligned}
    \label{eqn:B_S}
\end{equation}
with 
\begin{align}
    L^{m} &= [I- S^{m}_{m} {S^{m}_{m}}^{\dagger}]^{-\frac{1}{2}},
    \label{eqn:L}\\
    R^{m} &= [I- {S^{m}_{m}}^{\dagger} S_{m}^{m}]^{\frac{1}{2}}.
    \label{eqn:R}
\end{align}

An equivalent form of this relation is
\begin{equation}
    S^{m}(z) = [I+V^{m}(z){S^{m}_{m}}^{\dagger}]^{-1} [V^{m}(z)+S^{m}_{m}]\label{eqn:S_V}
\end{equation} 
with 
\begin{equation}
    V^{m}(z)= \frac{|z_{m}|(z_{m}-z)}{z_{m}(1-z_{m}^{*}z)} (L^{m})^{-1} B^{m}(z) (R^{m})^{-1}.
    \label{eqn:V_B}
\end{equation}

The relation between $B^{m}(z)$ and $S^{m+1}(z)$ reads
\begin{equation}
\begin{aligned}
    S^{m+1} (z) 
    & =[I- K^{m} {K^{m}}^{\dagger}]^{- \frac{1}{2}} [B^{m}(z)-K^{m}] \\
    &\cdot [I-{K^{m}}^{\dagger} B^{m}(z)]^{-1} [I-{K^{m}}^{\dagger}K^{m}]^{\frac{1}{2}},
\end{aligned}
\label{eqn:S_B}
\end{equation}
where $K^{m}$ is an arbitrary matrix with $||K^{m}||<1$.

The freedom in choosing arbitrary $K^{0}$, $K^{1}$, $\cdots$, $K^{N-1}$, along with $S^{N}(z)$, allows for a full coverage of all possible interpolants for $S(z)$. The seemingly undetermined quantities in the aforementioned equations are the values of $\set{S^{m}_{m}|m=0, 1,\cdots , N-1 }$. Starting from the given $\set{S^{0}_{n}| n=0, 1, \cdots , N-1}$, as $S^{m+1}_{n}$ are entirely determined by $S^{m}_{n}$ and $S^{n}_{n}$ by proceeding from Eq.~(\ref{eqn:B_S}) to Eq.~(\ref{eqn:S_B}), all $S^{m}_{m}$ can be acquired in the process of calculating $\set{S^{m}_{n}| m=0,1, \cdots N-1; n=m,m+1, \cdots ,N}$ from $m=0$ to $m=N-1$. After obtaining all $S^{m}_{m}$ and the corresponding $L^{m}$ and $R^{m}$, one can reverse the procedure, transitioning from Eq.~(\ref{eqn:S_B}) to Eq.~(\ref{eqn:S_V}), to compute $S^{m}(z)$ from $S^{m+1}(z)$ with a chosen $S^{N}(z)$ as the starting point, until the desired interpolant $S^{0}(z)$ is attained. Finally, an interpolant for $C(z)$ can be derived using Eq.~(\ref{eqn:C_S}), and by multiplying $-\mathrm{i}$ and substituting $z$ with $\tilde{z}$ through the M\"{o}bius transform, the interpolant for $G(\tilde{z})$ is obtained.

%  We take ${\tilde{z}_{n}} = {\mathrm{i} \omega{n}| n=0,1, \cdots , N-1}$ along with the corresponding $G(\mathrm{i} \omega_{n})$ as input, where $\omega_n = (2n+1)\pi/\beta$. Proceeding through $\tilde{F}=\mathrm{i} G$, Eqs.(\ref{eqn:Mobius}),(\ref{eqn:C_F}), and(\ref{eqn:S_C}), we obtain the constraints ${(z_n,S_{n})}$ for the interpolation problem for a Schur function $S(z)$. By setting $K^{m}=0$ for all $m$s, from Eq.(\ref{eqn:S_B}), we derive $ S^{m+1}(z) = B^{m} (z)$.

For the purpose of this work, we take $\{\tilde{z}_{n}\} = \{\mathrm{i} \omega_{n}| n=0,1, \cdots , N-1\}$ along with the corresponding $G(\mathrm{i} \omega_{n})$ as the input. Setting $\tilde{F}=\mathrm{i} G$, and with Eqs.~(\ref{eqn:Mobius}),~(\ref{eqn:C_F}), and~(\ref{eqn:S_C}), we obtain the constraints $\{(z_n,S_{n})\}$ for the interpolation problem of a Schur function $S(z)$. By taking $K^{m}=0$ for all $m$s, from Eq.(\ref{eqn:S_B}), we obtain $ S^{m+1}(z) = B^{m} (z)$
which leads to
\begin{equation}
      S^{m+1}_{n} 
       = \frac{z_{m}(1-z_{m}^{*}z_{n})}{|z_{m}|(z_{m}-z_{n})}  L^{m} [S^{m}_{n}-S^{m}_{m} ]  [I - {S_{m}^{m}}^{\dagger} S^{m}_{n}]^{-1} R^{m}
\end{equation}
according to Eq.~(\ref{eqn:B_S}). This equation offers a concrete way to calculate all $S^{m}_{n}$, starting from $S^{0}_{n}=S_{n}$. Subsequently, we can obtain $S^{m}_{m}$ and the corresponding $L^{m}$ and $R^{m}$. Assume the desired new Matsubara points for a different temperature $\beta^{\prime}$ are $\{\mathrm{i} \omega_{n}^{\prime}| n=0,1, \cdots , N-1\}$. Using the M\"{o}bius transform, the corresponding points in the domain of $S(z)$ are obtained as ${z^{\prime}_{n}}$. By setting $S^{N}(z)=I$, and using the obtained $\{S^{m}_{m}, L^{m}, R^{m}\}$, the values of ${S_{n'}\equiv S^{0} (z^{\prime}_{n})}$ can be determined by proceeding from Eq.~(\ref{eqn:S_B}) to Eq.~(\ref{eqn:S_V}). With the inverse Cayley transform in Eq.~(\ref{eqn:C_S}) and $G=-\mathrm{i} \tilde{F}$, the desired values of the causal interpolant ${G(\mathrm{i} \omega^{\prime}_n)}$ at ${\mathrm{i} \omega^{\prime}_n}$ are obtained as output.
%\textcolor{magenta}{This previously commented out paragph seems to be inevitable though it has overlaps with previous paragphs, it mentions in practice: we need take $K^{m}=0$; take $S^N(z)=I$; evaluate at new frequencies points. Eq. 11 also makes the algorithm more clear and is used in Fig. 1.}

\section{Implementation}\label{sec:impl}
We provide a straightforward  pedagogical implementation of the algorithm  as supplement to this paper \footnote{See Supplemental Material at \href{https://github.com/CQMP/pyCaratheodory}{https://github.com/CQMP/pyCaratheodory} for Python implementation of the algorithm, documentation, and examples.}. The code is written in the \texttt{python} programming language, using no dependencies other than \texttt{numpy}. Unlike related real-frequency analytic continuation codes \cite{fei2021a,fei2021b}, we find that an implementation in standard double precision was sufficient to perform all calculations, and that the temperature extrapolation of noisy Monte Carlo data presented no  difficulties.

The implementation requires a Green's function (or, equivalently, self-energy or cumulant) $G[n,i,j]=G_{ij}(\mathrm{i}\omega_n)$ as a three-dimensional tensor and $w[n]=\mathrm{i}\omega_{n}$ as a one-dimensional vector. After reading the data, all preprocessing steps are performed until $\{S^{m}_{m}, L^{m}, R^{m}\}$ are obtained and stored, as explained in Sec~\ref{sec:method}. 

The code also requires a set of frequencies $wp[n]=\mathrm{i}\omega^{\prime}_{n}$, provided as a one-dimensional vector. These frequencies should correspond to the Matsubara points at the new temperature $\beta^{\prime}$. Using this data, the code evaluates the interpolated Green's function at the new Matsubara points and returns $G[n',i,j] = G_{ij}(\mathrm{i} \omega^{\prime}_{n})$ as a three-dimensional tensor.

The supplemental materials contain detailed instructions and usage examples for the code. 

%\begin{figure*}[tbh]
%    \centering
%    \includegraphics[width=1\linewidth]{dca_16kpts_sigma_diff.pdf}
%    \caption{The deviation of the imaginary part of interpolated $\Sigma(k_x,k_y)$ away from the target value in the 2D Hubbard model DCA calculation at first three Matsubara frequencies %($U=6$) as a function of the temperatures of starting points. The target temperature is $\beta=20$.}
%    \label{fig:temp}
%\end{figure*}

\section{Results}\label{sec:results} 
We showcase results for the temperature extrapolation technique applied to a range of typical self-consistent finite-temperature methods. Sec.~\ref{subsec:extrapol} investigates the accuracy and convergence properties of the extrapolation technique. Sec.~\ref{subsec:convergence} examines the convergence acceleration resulting from an enhanced starting point provided by the extrapolation technique. This study is conducted specifically within the context of single-site DMFT and cluster DMFT calculations for the Hubbard model, as well as the self-consistent GW approximation for nickel oxide (NiO) real material calculations. 
Sec.~\ref{subsec:comparison} presents a comparison between the method introduced in this paper and other methods for generating starting points. Sec.~\ref{subsec:HeatCool} demonstrates the occurrence of hysteresis during heating and cooling processes at a first-order phase transition, utilizing the extrapolation technique.

\subsection{Temperature extrapolation}\label{subsec:extrapol}
\begin{figure}[tb]
    \centering
    \includegraphics[width=\columnwidth]{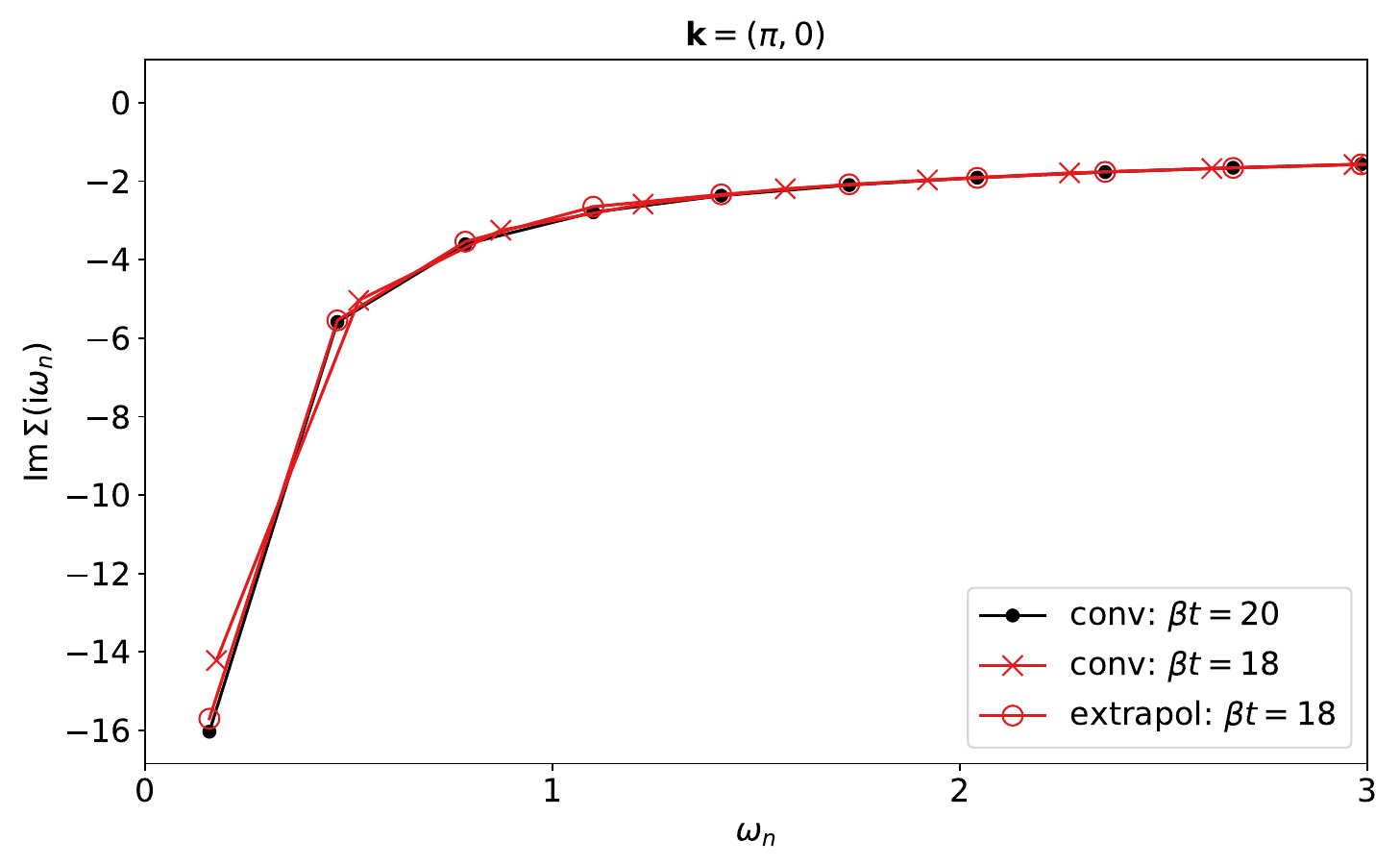}
    \caption{Imaginary part of a cluster DMFT \cite{maier2005a} self-energy, $\operatorname{Im} \Sigma(\mathrm{i}\omega_n)$, as a function of $\omega_n$ for the 2D Hubbard model at the antinodal point $(\pi,0)$ with $U=6 t$ (see  text for calculation setup). Converged data at $\beta t=18$ (red crosses) and $\beta t=20$ (black dots) along with  temperature extrapolation (red circle) from $\beta t=18$ to $\beta t=20$. The labels of the extrapolation results indicate the initial inverse temperature used for extrapolation. } 
    \label{fig:HubbardTExtrapol}
\end{figure}
\begin{figure}[htb]
    \centering
    \includegraphics[width=\columnwidth]{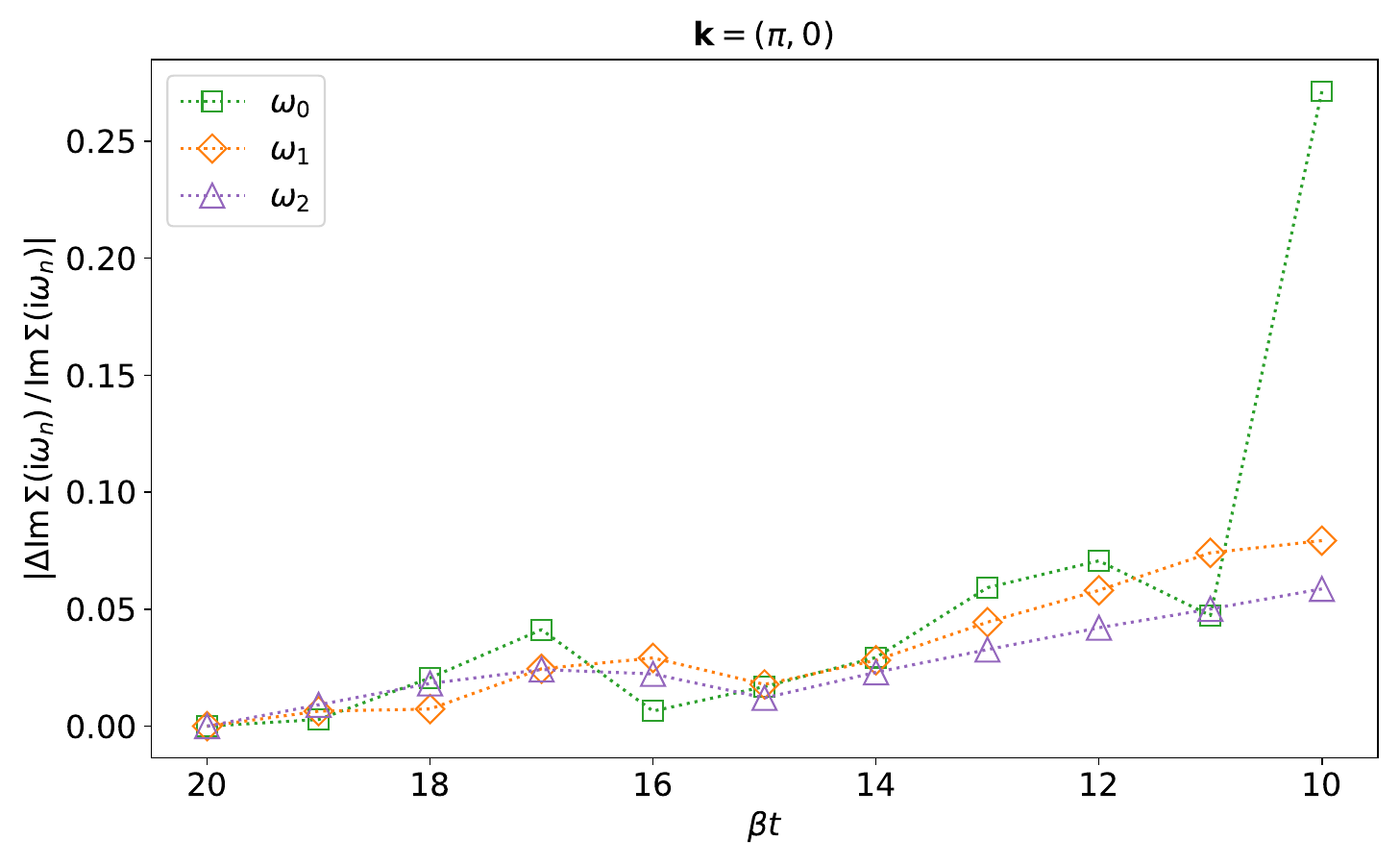}
    \caption{Difference $|\Delta \text{Im}\Sigma(\mathrm{i}\omega_{n})/\text{Im} \Sigma(\mathrm{i}\omega_{n})|$ between the converged self-energy at $\beta t=20$ and the extrapolation from the converged self-energy at higher temperature, shown as a function of the extrapolation inverse temperature for the three lowest Matsubara frequencies $\omega_n$, with $n=0, 1, 2$ (see  text for calculation setup).}
    \label{fig:temp}
\end{figure}
%We present an example of temperature extrapolation in a cluster DMFT calculation, as depicted in Fig.~\ref{fig:HubbardTExtrapol}. Specifically, we show the self-energy at the so-called antinodal point, $(\pi,0)$, for a simulation of the two-dimensional Hubbard model \cite{Qin22} on a 16-site ($4\times 4$) cluster, employing a continuous-time auxiliary field \cite{Gull08_ctaux,Gull11} impurity solver.
%The system is solved at half filling, in the paramagnetic state, without next-nearest neighbor hopping, and with an interaction strength of $U=6t$ which corresponds to a Mott insulating regime. The self-energy in this regime exhibits a strong temperature dependence, and various phases such as metallic, insulating, superconducting, and pseudo-gapped phases meet in close proximity \cite{Werner09,Gull09,Gull10}. We refer the reader to the extensive literature on this system (see reviews \cite{maier2005a,Qin22} and references therein) for a comprehensive discussion of the physics of this self-energy.
In this section, we provide an illustration of temperature extrapolation in a cluster DMFT  \cite{maier2005a} calculation, as depicted in Figure~\ref{fig:HubbardTExtrapol}. Specifically, we present the self-energy at the antinodal point, $(\pi,0)$, for a simulation of the two-dimensional Hubbard model \cite{qin2022a}. The simulation was performed on a 16-site ($4\times 4$) cluster, and the continuous-time auxiliary field impurity solver~\cite{gull2008b,gull2011b} was employed. The system was solved at half-filling, in the paramagnetic state, with no next-nearest neighbor hopping, and at an interaction strength of $U=6t$, which corresponds to a Mott insulating regime. The self-energy in this regime exhibits a strong temperature dependence, and various phases such as metallic, insulating, superconducting, and pseudo-gapped phases are in close proximity \cite{werner2009,gull2009,gull2010}. For a comprehensive discussion of the physics of this self-energy, we refer the reader to the extensive literature on this system, including the reviews \cite{maier2005a,qin2022a} and references therein.

Fig.~\ref{fig:HubbardTExtrapol} presents the fully converged, self-consistent imaginary part of the self-energy, $\Sigma(\mathrm{i} \omega_n)$, at $\beta t=18$ (red crosses) and at $\beta t=20$ (black dots). Red circles illustrate the extrapolation of the $\beta=18 t$ self-energy to $\beta=20 t$ using the Carath\'{e}odory formalism. The extrapolated data aligns with the converged points for all frequency points, except for the lowest one. The discrepancy between converged data and extrapolation data at $\beta=20 t$ serves as an indicator of additional correlations emerging in the system as it cools down.

Fig.~\ref{fig:temp} demonstrates the deviation of the extrapolated self-energy at the three lowest Matsubara frequencies at $\beta t=20$. The extrapolation is conducted using converged data at $\beta t=10,11, ..., 19$. Notably, while the lowest frequency of the self-energy exhibits significant deviations when extrapolated by a factor of two from $\beta t=10$ to $\beta t=20$, these deviations rapidly diminish if the extrapolation interval (the difference between the initial temperature and targeted temperature) is reduced. 

Figs.~\ref{fig:HubbardTExtrapol} and~\ref{fig:temp} demonstrate that the Carath\'{e}odory structure of many-body functions can be effectively utilized to extrapolate data in temperature.

% As shown in Fig.~\ref{fig:temp}, the majority of the deviation is caused by the emergence of additional correlations as temperature is lowered. \textcolor{magenta}{Should we make it more clear here by adding ``the lowest frequency has the largest extrapolation error on average''}

\begin{figure}[tb]
    \centering
    \includegraphics[width=\columnwidth]{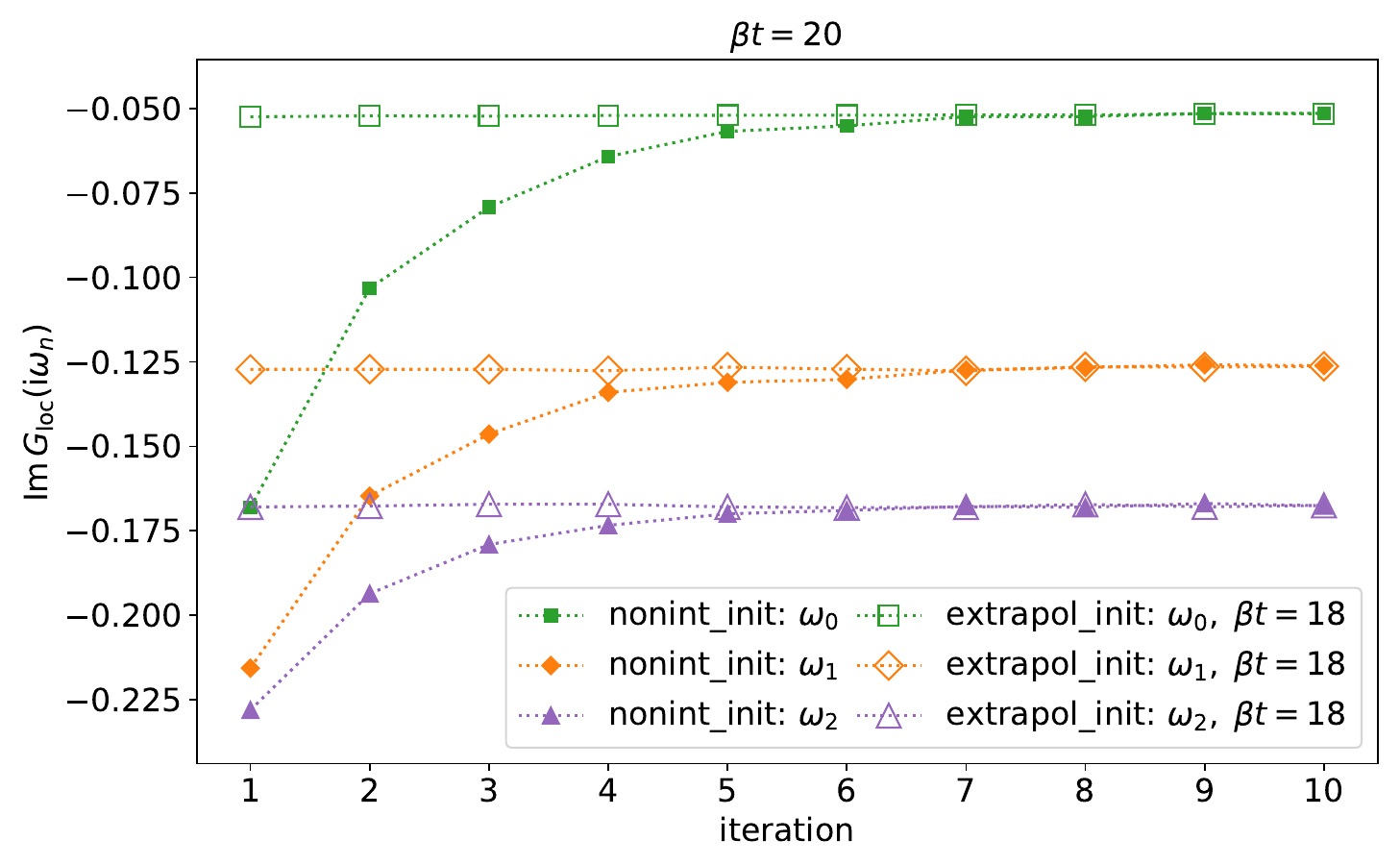}
    \caption{Cluster DMFT convergence of the imaginary part of the local Green's function $G_{\text{loc}}$ as a function of iteration for the three lowest Matsubara frequencies at $\beta t=20$ (see text for calculation setup). Convergence to the fixed point is shown for the non-interacting starting point (filled symbols) and the starting point extrapolated from the converged solution at $\beta t=18$ (open symbols). The labels of the extrapolation results indicate the initial inverse temperature used for extrapolation. 
     \label{fig:ConvergenceWithStartingPoint}
     }
\end{figure}

\subsection{Starting point and Convergence}\label{subsec:convergence}
\subsubsection{Convergence of the extrapolated starting point away from phase transitions}
A temperature extrapolation of this type can be used to improve the starting point for the convergence of self-consistent many-body methods, such as DMFT~\cite{metzner1989,georges1992,georges1996a} and its cluster extensions~\cite{hettler1998, lichtenstein2000a,kotliar2001a,maier2005a}. The DMFT equations are solved using a fixed-point iteration scheme, in which an initial estimate of the self-energy (commonly chosen as $\Sigma(i\omega_n) \equiv 0$) serves as an initial guess for the iteration. The equations typically converge rapidly if the initial starting point of the iteration is near the stationary point. Since the computational effort is proportional to the number of iterations required, a starting point closer to the stationary point reduces the calculation time.

Fig.~\ref{fig:ConvergenceWithStartingPoint} displays the imaginary part of the lowest three frequencies of the local Green's function, $G_{\text{loc}}(\mathrm{i} \omega_{n})$, for a typical cluster DMFT calculation away from phase transitions, using non-interacting starting points (filled symbols) and starting points derived from temperature extrapolation (open symbols), as a function of iteration. The calculation setup is identical to the one in Sec.~\ref{subsec:extrapol} (a 16-site cluster in the paramagnetic state with $\beta t=20$, $U=6t$, and at half filling). A naive non-interacting starting point converges in approximately seven iterations. Moreover, a few additional iterations are necessary to confirm that convergence has been achieved. In contrast, a starting point generated by extrapolating from higher to lower temperatures almost immediately yields a converged result.

\subsubsection{DMFT convergence near a phase transition}

\begin{figure}[tb]
    \centering
    \includegraphics[width=\columnwidth]{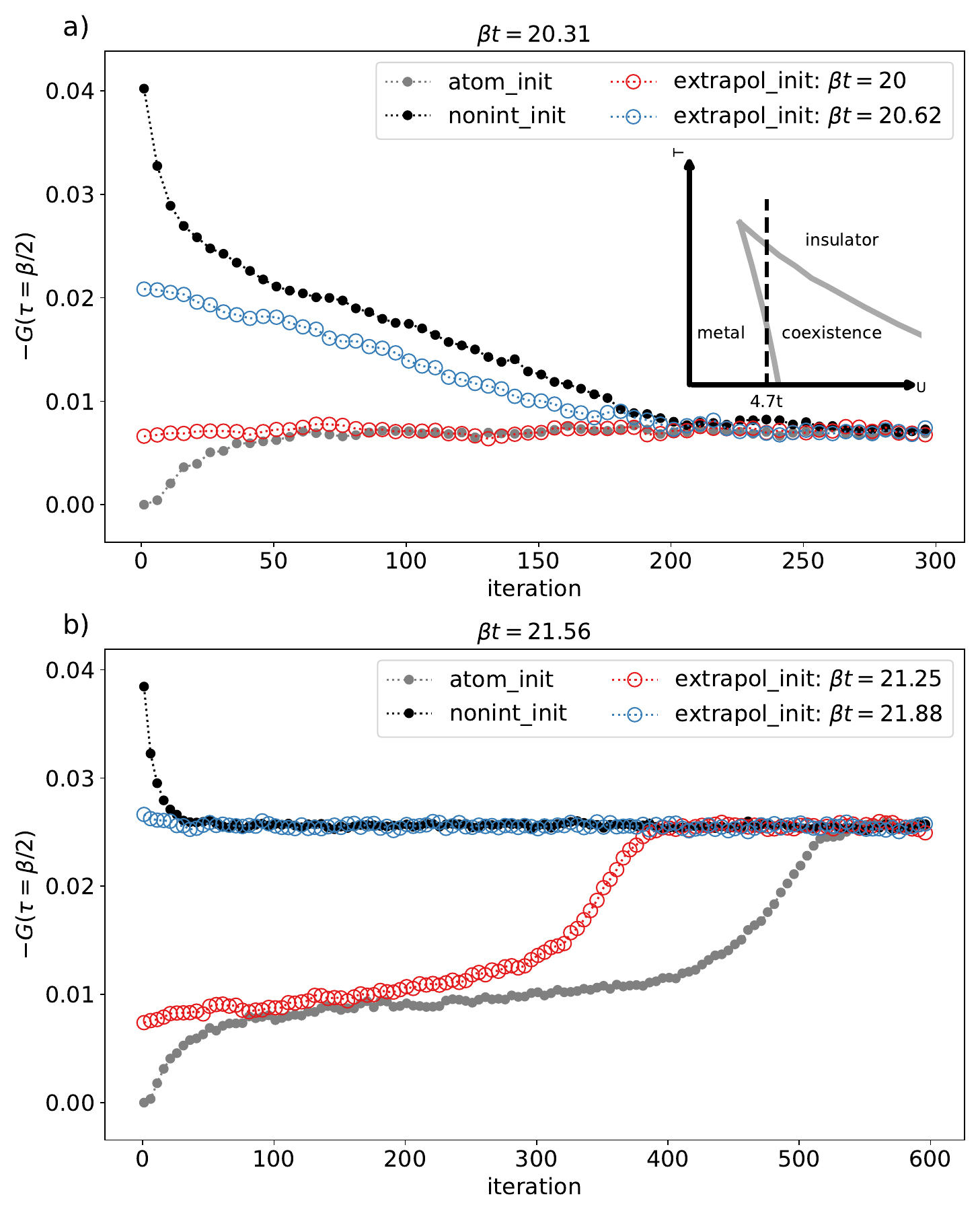}\\
    \caption{DMFT convergence of $-G(\beta/2)$ on a half-filled inifinite-coordination Bethe lattice with $U=4.7t$ and bandwidth $W=4t$ as a function of iteration.  The top and lower panels display the insulating ($\beta t=20.31$) and metallic ($\beta t=21.56$) phases, respectively. Atomic starting point: gray dots. Non-interacting starting point: black dots. Extrapolated starting points: red circles and blue circles. Inset: phase diagram adapted from Ref.~\cite{Bluemer03}. The labels of the extrapolation results indicate the initial inverse temperature used for extrapolation. Iteration points with an interval of 5 are shown for better visibility.}
    \label{fig:Bethe}
\end{figure}

Self-consistent many-body methods are notoriously slow to converge in the vicinity of phase transitions. This can be illustrated at the example of a single-site DMFT calculation on an infinite coordination number Bethe lattice with bandwidth $W=4t$.  The single-site DMFT calculation exhibits a first-order phase transition between a paramagnetic metal at weak interaction and low temperature and a paramagnetic Mott insulator at large interaction and higher temperature \cite{georges1992,jarrell1992,rozenberg1992,zhang1993,rozenberg1999,bulla1999,Bluemer03}, as depicted in the inset of Fig.~\ref{fig:Bethe}. At an interaction strength $U=4.7t$, the system is insulating at high temperature, metallic at low temperature, and both metallic and insulating solutions coexist in an intermediate temperature regime. We show the convergence of the imaginary-time Green's function in the middle of the imaginary time interval, $-G(\beta/2)$, in Fig.~\ref{fig:Bethe}, which is related to the spectral density at the Fermi surface in the low-temperature limit as $A(\omega=0)\sim-\beta G(\beta/2)$, at temperatures close to the two boundaries of the coexistence region.

At a temperature of $\beta t=20.31$, the system exhibits Mott insulating behavior. The upper panel of Fig.~\ref{fig:Bethe} shows that hundreds of iterations are required to achieve convergence when initiating from the non-interacting limit (black dots). Similarly, starting with an extrapolation from a lower temperature metallic solution in the coexistence region (blue circles) results in slow convergence toward the insulating fixed point. Conversely, the extrapolation from a higher temperature insulating phase converges rapidly (red circles). A starting point derived from the atomic limit also leads to relatively fast convergence (gray dots).

Conversely, at a temperature of $\beta t=21.56$ (Fig.~\ref{fig:Bethe} lower panel), the system is in a metallic state. Convergence from an `insulating' starting point requires hundreds of iterations to reach convergence (red circles for the starting point extrapolated from a higher temperature insulating solution, and gray dots for the atomic limit starting point). Convergence from a metallic solution, like the non-interacting limit solution (black dots), is substantially faster, and a starting point extrapolated from a converged metallic solution at lower temperature leads to even faster convergence (blue circles).

Fig.~\ref{fig:Bethe} therefore shows that even though all starting points, including atomic, non-interacting, and extrapolated, converge to the identical solution, extrapolation efficiently accelerates convergence, provided that the extrapolated solution is in the same phase as the solution at the target temperature. For extrapolation originating from a different phase (at the boundary of phase transitions, solutions with distinct properties become possible for two neighboring temperatures), the extrapolated starting points still perform better than the simplistic starting points corresponding to the `wrong' phase, specifically, the non-interacting solution or the atomic solution.

\begin{figure}[tb]
    \centering
    \includegraphics[width=1\linewidth]{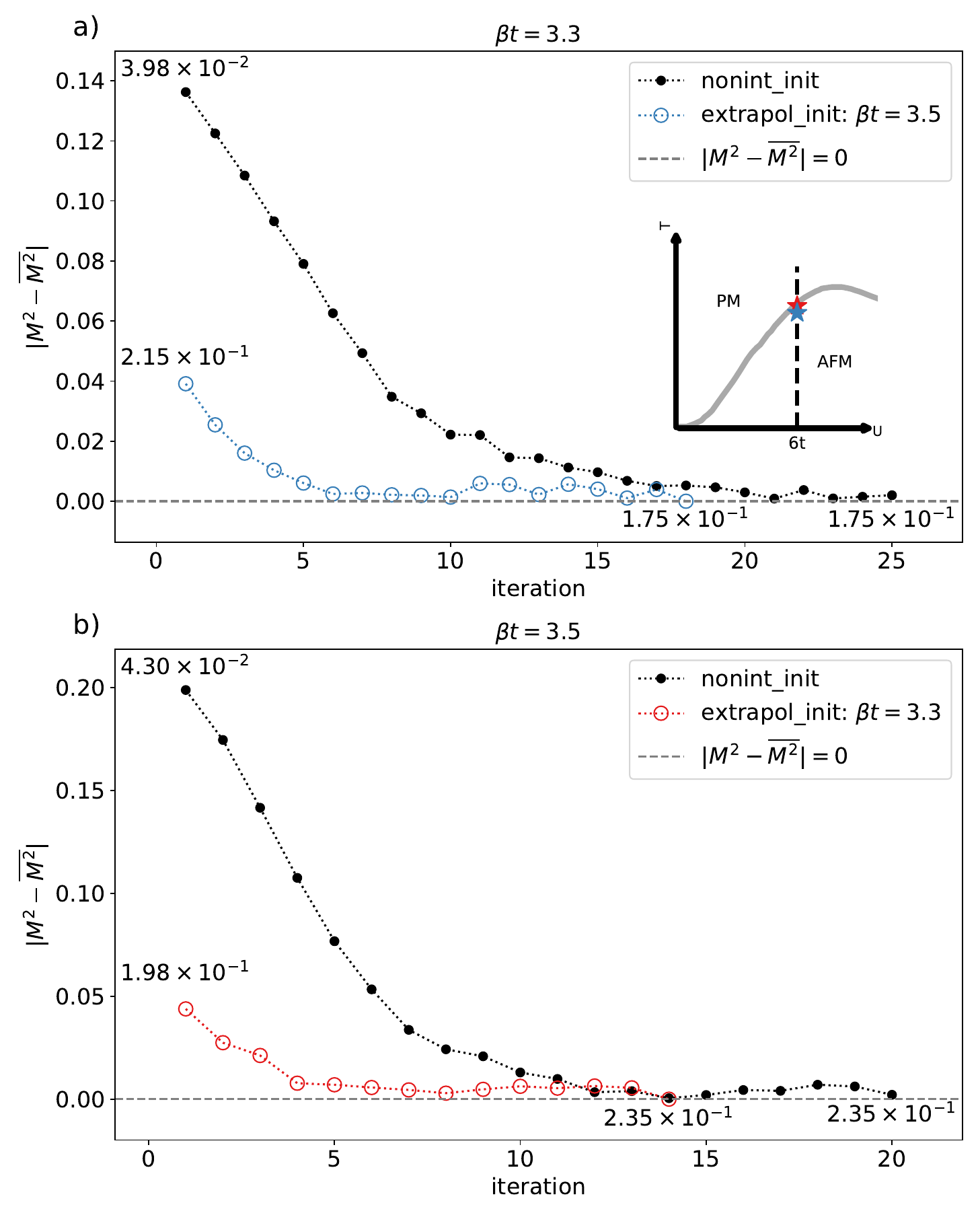}
    \caption{Cluster DMFT convergence of the squared magnetic moment on the half-filled 3D Hubbard model (using a 34-site cluster with a doubled unit cell) at $U=6t$ and $\beta t=3.3$ (upper panel) and $3.5$ (lower panel) as a function of iteration. Dots: non-interacting starting point. Circles: temperature extrapolation. Inset: phase diagram adapted from Ref.~\cite{staudt2000a}. $M^2$ at the first and last iteration indicated by numbers. $\overline{M^2}$ corresponds to the values at the last iteration of red or blue curves, which serves as a reference for converged values. The labels of the extrapolation results indicate the initial inverse temperature used for extrapolation. }
    \label{fig:3D-Hubbard}
\end{figure}
\subsubsection{Cluster DMFT convergence near a phase transition}
At second order phase transitions, critical slowing down, rather than coexistence and hysteresis, is expected. This phenomenon leads to a slow convergence of the fixed point iteration. We illustrate this at the example of a half-filled three-dimensional Hubbard model treated within cluster DMFT~\cite{kent2005,fuchs2011,fuchs2011a}, calculated on a 34-site cluster with a doubled unit cell. This model exhibits a phase transition between a paramagnetic (PM) state at high temperature and an antiferromagnetic (AFM) insulator at low temperature, as depicted in the inset of Fig.~\ref{fig:3D-Hubbard}. The model has been studied extensively within the context of ultracold atomic gases. In Fig.~\ref{fig:3D-Hubbard}, we examine two points, $\beta t=3.3$ and $\beta t=3.5$, in the ordered phase at $U = 6t$, near the AFM phase transition (the critical point is at about $\beta t=2.7$). A starting point from temperature extrapolation cannot completely overcome the slowing down (since the self-energy itself exhibits strong temperature dependence) but does lead to an acceleration of the convergence by at least a factor of two, as illustrated by the convergence of the squared magnetic moment shown in Fig.~\ref{fig:3D-Hubbard}.

\begin{figure*}[tb]
    \centering
    \includegraphics[width=1\linewidth]{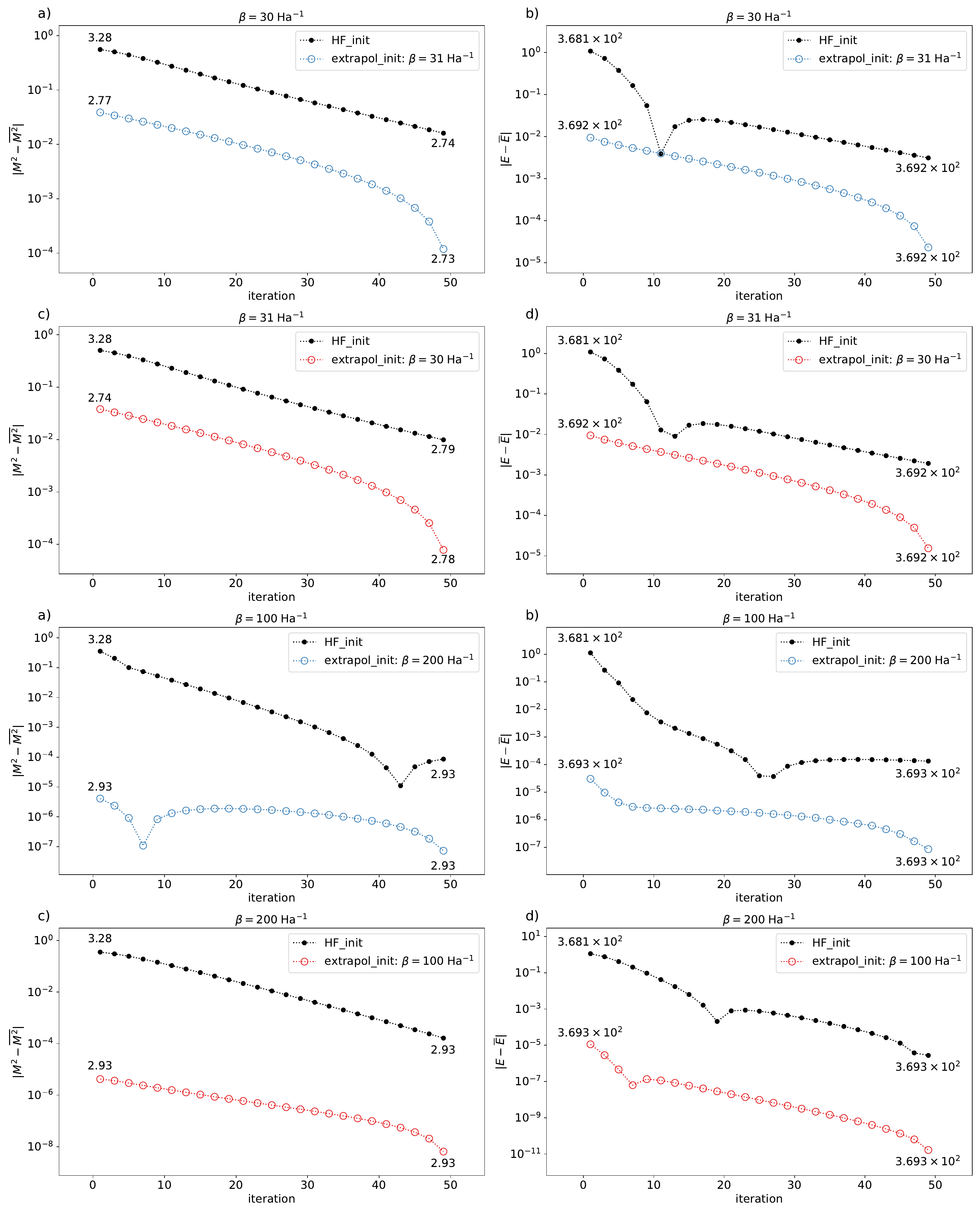}
    \caption{GW convergence of the squared magnetic moment (left) and energy (right) near the AFM phase transition of NiO at $\beta =30\;\mathrm{Ha}^{-1}$ (upper panel) and $\beta =31\;\mathrm{Ha}^{-1}$ (lower panel) as a function of iteration. Calculations with unrestricted Hartree-Fock self-energy starting point are represented by black dots, while calculations with temperature-extrapolated self-energy are denoted by red circles (indicating extrapolation from higher temperature solutions) or blue circles (indicating extrapolation from lower temperature solutions). The Hartree atomic units are used.  The labels of the extrapolation results indicate the initial inverse temperature used for extrapolation. Iteration points with an interval of 2 are shown for better visibility. The values of $M^2$ and $E$ at the first and last iterations are indicated by the numbers. The values of $\overline{M^2}$ and $\overline{E}$ are determined as the values at the last iteration of red or blue curves (The symbols for the last iteration are not shown.). }
    \label{fig:NiONearPT}
\end{figure*}
% \begin{figure*}[tb]
%     \centering
%     \includegraphics[trim=0 0 0 30.5cm, clip,width=1\linewidth]{GW_NiO.pdf}
%     \caption{GW convergence of the squared magnetic moment (left) and energy (right) deep within the AFM phase of NiO at $\beta =100\;\mathrm{Ha}^{-1}$ (upper panel) and $\beta =200\;\mathrm{Ha}^{-1}$ (lower panel) as a function of iteration. Calculations with unrestricted Hartree-Fock self-energy starting point are represented by black dots, while calculations with temperature-extrapolated self-energy are denoted by red circles (indicating extrapolation from higher temperature solutions) or blue circles (indicating extrapolation from lower temperature solutions). The values of $M^2$ and $E$ at the first and last iterations are indicated by the numbers. The values of $\overline{M^2}$ and $\overline{E}$ are determined as the values at the last iteration of red or blue circles. The Hartree atomic units are used.  The labels of the extrapolation results indicate the initial inverse temperature used for extrapolation.}
%     \label{fig:NiOLowT}
% \end{figure*}

\subsubsection{GW convergence of realistic many-body simulations}
Finally, we turn to realistic simulations within a weak-coupling framework. We show examples for periodic solid NiO treated within the so-called GW approximation \cite{hedin1965,aryasetiawan1998,kutepov2009,kutepov2020,yeh2022a,yeh2022b}. The GW approximation takes into account screening processes via a renormalized, frequency-dependent interaction, but neglects the second-order exchange diagram. It is therefore mostly used for weakly correlated systems such as semiconductors. In the calculation for NiO, a 4x4x4 cluster with a doubled unit cell along the $[1,1,1]$ direction and a fixed lattice constant $a=4.1705$ \AA~\cite{bartel1971} is utilized. The \textit{gth-dzvp-molopt-sr} basis~\cite{vandevondele2007} along with the \textit{gth-pbe} pseudopotential~\cite{goedecker1996} is employed. For the density fitting of Coulomb integrals, the \textit{def2-svp-ri} basis set is chosen as the auxiliary basis~\cite{hattig2005}. Finite-size errors in the GW exchange diagram are corrected using the Ewald probe-charge approach~\cite{paier2005,sundararaman2013}. The Coulomb integrals and non-interacting matrix elements are obtained from PYSCF~\cite{sun2018}. In order to decrease the number of frequencies utilized in the computation, the IR grid~\cite{shinaoka2017} is employed. A comprehensive description of the methods and implementation, in conjunction with the computational setup for evaluating NiO, is extensively detailed in Refs.~\cite{iskakov2020a,yeh2022a}. 

When applied to the antiferromagnetic materials NiO, the GW method shows a continuous transition to an ordered state with a non-zero staggered magnetization at low temperature. Within the GW framework, the transition temperature is situated near $\beta=25\;\mathrm{Ha}^{-1}$. As illustrated in Fig.~\ref{fig:NiONearPT}, the convergence to this ordered state, initiated from an unrestricted Hartree-Fock solution (black dots), is relatively slow. This is evident in the convergence of both the squared magnetic moment for Ni and the total energy. However, when starting from extrapolated starting points (blue or red circles), the convergence occurs at a significantly faster pace, with an initial starting point akin to iteration 30 of the Hartree-Fock-initialized convergence.

% At considerably lower temperatures, deep within the AFM phase, the extrapolation of the self-energy results in a starting point that is essentially exact because the self-energy is only weakly temperature-dependent, similar to the 2D Hubbard model case shown at the beginning of Sec.~\ref{subsec:convergence}. Fig.~\ref{fig:NiOLowT} displays the temperature extrapolation from $\beta t=100$ to $\beta t=200$, and vice versa, in comparison with the Hartree-Fock starting points.

\subsection{Comparison with other methods}\label{subsec:comparison}
\begin{figure}[htb]
    \centering
    \includegraphics[width=1\linewidth]{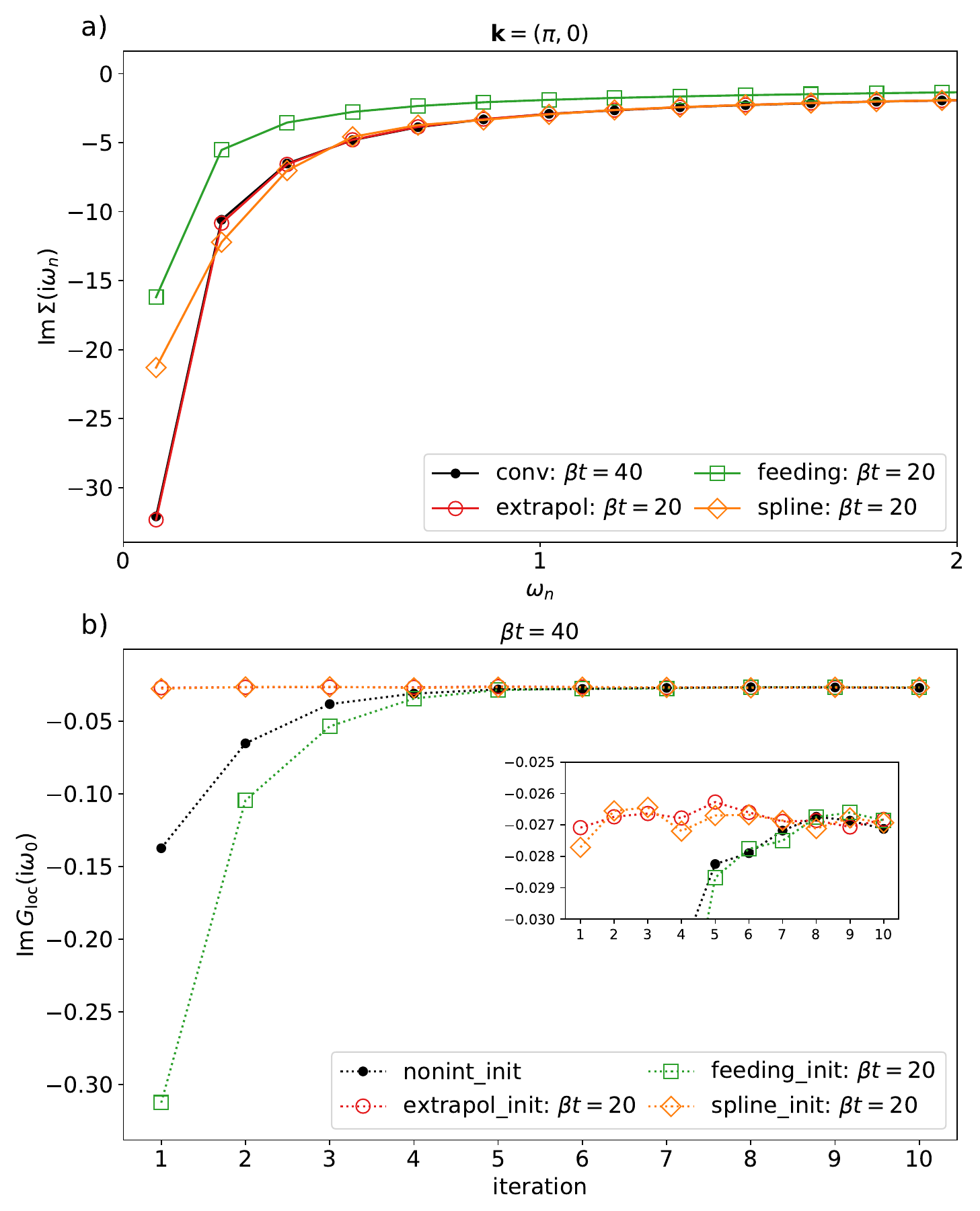}
    \caption{Comparison for different acceleration methods for a 16-site cluster DMFT calculation for the Hubbard model at $U=6t$.  All temperature extrapolations are performed from $\beta t = 20$ to $\beta t = 40$. The labels of the
    extrapolation results indicate the initial inverse temperature
    used for extrapolation. (a)  The imaginary part of the self-energy, $\operatorname{Im} \Sigma (\mathrm{i} \omega_{n})$, at the antinodal point ($\pi$, 0) as a function of $\omega_{n}$. Converged data at $\beta t=40$: black dots; extrapolated data via Carathéodory formalism: red circles; extrapolated data via temperature feeding: green squares;  extrapolated data via cubic spline interpolation: orange diamonds. (b) The imaginary part of the local Green’s function $G_{\mathrm{loc}}$ at the lowest Matsubara frequencies $\omega_{0}$ as a function of iteration.  Noninteracting starting point: black dots; Carathéodory starting point: red circles;  temperature feeding starting point: green squares; cubic spline starting point: orange diamonds. }
    \label{DCA_16kpts_comparsion}
\end{figure} 
\begin{figure}[htb]
    \centering
    \includegraphics[width=1\linewidth]{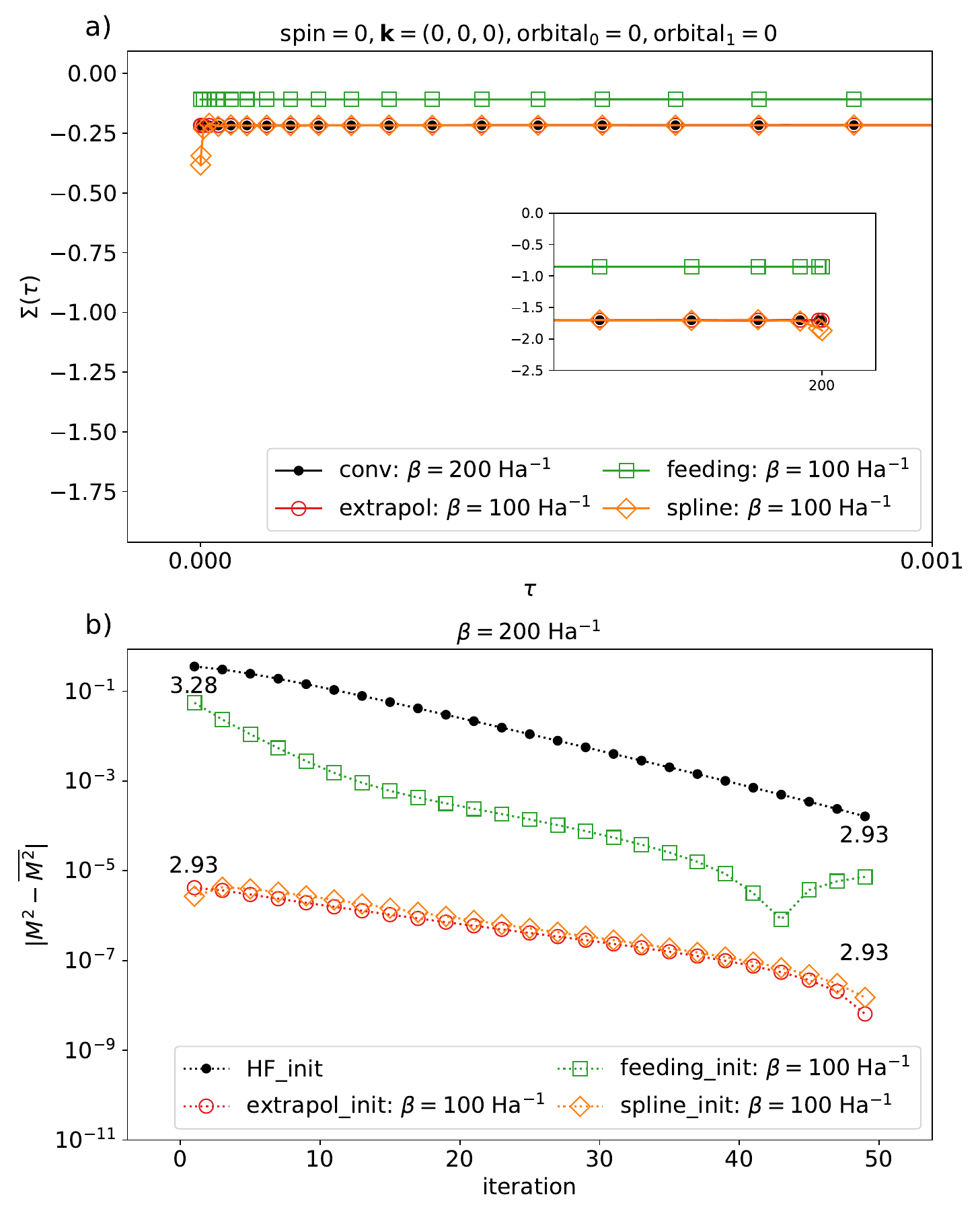}
    \caption{Comparison for different acceleration methods for a self-consistent GW calculation for NiO.  All temperature extrapolations are performed from $\beta = 100 \; \mathrm{Ha}^{-1}$ to $\beta = 200 \; \mathrm{Ha}^{-1}$. The labels of the
    extrapolation results indicate the initial inverse temperature
    used for extrapolation. (a) Imaginary time self-energy, $\operatorname{Im} \Sigma (\tau)$, with fixed spin, momentum, and orbital indices. Converged data at $\beta =200 \; \mathrm{Ha}^{-1}$: black dots; extrapolated data via Carathéodory formalism: red circles; extrapolated data via temperature feeding: green squares;  extrapolated data via cubic spline interpolation: orange diamonds. (b) Convergence of the squared magnetic moment as a function of iteration.  Hartree-Fock starting point: black dots; Carathéodory starting point: red circles;  temperature feeding starting point: green squares; cubic spline starting point: orange diamonds. Iteration points with an interval of 2 are shown for better visibility.  The values of $M^2$ at the first and last iterations of the red and black curves are indicated by the numbers. The
    value of $\overline{M^2}$ is determined as the value at the last iteration of the red curve (The symbol for the last iteration is not shown.).}
    \label{GW_NiO_comparsion}
\end{figure} 

By assuming the causal matrix-valued functions $G(\tilde{z})$ at nearby temperatures having a close analytical structure, we introduce the temperature extrapolation for the Matsubara functions by evaluating the interpolant of $G(\tilde{z})$ of one temperature $\beta$ on the Matsubara grid for another temperature $\beta^{\prime}$.
Besides the Carathéodory formalism, there are other interpolation methods available for the same purpose, warranting comparative analysis. We compare the Carathéodory formalism with a common alternative: cubic spline. For cubic spline, additional constraints are needed at the boundaries to fix all the coefficients for cubic polynomials. After obtaining the spline, one could use the boundary cubic polynomial for extrapolating values outside the original scope or force the extrapolated segment to maintain the derivative values at the boundary (for instance, performing linear extrapolation if we set the boundary's second derivative to zero).  For the Matsubara functions under study, we find the `not-a-knot spline'~\cite{burden2016}, which requires that the first two cubic polynomials at the boundary be identical, to be superior to other boundary constraints. Therefore, we only show this type of cubic spline for comparison, labeled as `spline'. In addition, we also compare with a commonly used technique to accelerate temperature scanning,  where the result at $\beta$ is directly used as a starting point for $\beta^{\prime}$  despite the mismatching of the Matsubara grids. This approach is referred to as `temperature-feeding' here and is labeled as `feeding'. Given that the challenging part of simulations often manifests at lower temperatures, this section primarily focuses on the cooling process. Furthermore, as $\beta'$ approaches $\beta$, all the discussed methods should provide reasonably good starting points and converge more rapidly than simpler methods, such as the noninteracting and Hartree-Fock starting points. Hence, we engage with a relatively large temperature difference here, setting $\beta^{\prime}$ to $2\beta$.

% \textcolor{red}{The additional interpolation techniques for comparison are in two catagories. One is the commonly used naive technique where the higher temperature results are directly used as a starting point for the next lower temperature step despite the mismatching of the Matsubara grids. We called it ``temperature-feeding" here. Another type of interpolation schemes employ more refined polynomial interpolation techniques with additional extrapolation to reach lower Matsubara frequencies. Here we focus on the most common and reliable one: cubic spline interpolation. To reach lower Matsubara frequencies, additional constraints are needed at the boundaries. Here we studies the following boundaries conditions: ``not-a-knot'': the first and second segment at a curve end are the same polynomial; ``clamped-bc'': the first derivative at curves ends are zero; ``natural-bc'': the second derivative at curve ends are zero; ``clamped'': forcing the extrapolated parts to have the same values as curve ends; ``natural'': forcing the extrapolated parts to be linear.}

Figure \ref{DCA_16kpts_comparsion} utilizes the identical 16-site cluster DMFT calculation setup for the Hubbard model as detailed in Sec.~\ref{subsec:convergence}, albeit this instance involves an extrapolation from $\beta t=20$ to $\beta t =40$.   In Figure \ref{DCA_16kpts_comparsion} (a), the imaginary part of self-energy at the antinodal point is displayed. The extrapolated self-energy via the Carathéodory formalism (red circles) closely aligns with the fully converged self-energy (black dots) at all Matsubara frequencies. In contrast, extrapolated self-energy given by the cubic spline (orange diamonds) exhibits mismatches at the initial few low Matsubara frequencies. As anticipated, the temperature-feeding method demonstrates discrepancies across all Matsubara frequencies. Figure \ref{DCA_16kpts_comparsion} (b) illustrates the different methods' convergence behaviour for the imaginary part of the local Green's function. Both the starting points of the cubic spline and Carathéodory methods display immediate convergence within the first two iterations, whereas temperature feeding worsens convergence compared to the noninteracting starting point. As observed in the inset, the advantages of the Carathéodory method's starting point over cubic spline's starting is about the same magnitude of Monte Carlo errors, which obscures the distinction in convergence between the two methods.

Figure \ref{GW_NiO_comparsion} employs the same self-consistent GW calculation setup for NiO as detailed in Sec.~\ref{subsec:convergence}, but in this case, we extrapolate from $\beta =100 \;\mathrm{Ha}^{-1}$ to $\beta =200 \;\mathrm{Ha}^{-1}$.   Figure \ref{GW_NiO_comparsion} (a) presents the imaginary time self-energy with fixed spin and orbital indices at the $\Gamma$ point. This imaginary time self-energy is the self-consistently converged quantity utilized in our GW code. For the extrapolation stage, we carry out the extrapolation in frequency space and then transform back to the imaginary time space on the IR grid~\cite{shinaoka2017}. The Carathéodory method (red circles) gives nearly perfect extrapolation in comparison with the fully converged self-energy, while the cubic spline (orange diamonds) exhibits discrepancies at both $\tau=0$ and $\tau=\beta$ (see inset). The temperature-feeding method, as anticipated, shows discrepancies over the entire time grid. We observe a similar behavior when performing extrapolation for the time-ordered Green's function $G(\tau)$, where the analytical properties $G_{ij}(0^{+})+G_{ij}(\beta^{-})=-\delta_{ij}$ (with fixed spin and momentum indices) are significantly violated by the starting points derived from the cubic spline and temperature feeding methods. In contrast, these properties are preserved to a high degree of precision for the starting point derived from the Carathéodory method. Figure \ref{GW_NiO_comparsion} (b) demonstrates the convergence behavior of different methods for the squared magnetic moment. Both the starting points of the cubic spline and Carathéodory methods exhibit superior convergence behavior compared to the Hartree-Fock starting point and the temperature feeding starting point, while the temperature feeding starting point outperforms the Hartree-Fock starting point. In this scenario, it can be observed that the Carathéodory starting point converges faster than the cubic spline starting point by one or two iterations (the interval between points shown in Fig.~\ref{GW_NiO_comparsion} (b) is two).

    From the two test cases detailed above, we note that at relatively large temperature differences, temperature feeding can either accelerate or decelerate convergence. Both cubic spline interpolation and Carathéodory formalism, when used for extrapolation, are superior choices for the starting points, with the Carathéodory formalism demonstrating a slight advantage. 
    We  stress that systems with a strongly divergent self-energy will likely exhibit more evident superiority of the Caratheodory method in comparison to cubic spline interpolation.
    We also emphasize that, while cubic spline interpolation proves effective in the specific example we examined, its combination with IR coefficients results in a substantial loss of accuracy and likely failure of a self-consistent calculation.
    % As the two test cases suggest, the optimal test case to highlight the difference between cubic spline and Carathéodory convergence would be one where the extrapolated quantity tends to diverge near the zero frequency, and the calculation is executed with minimal numerical fluctuation.
    % Although the GW method does not experience severe numerical fluctuations, the method's weak-coupling nature tends to prevent a divergent self-energy. We believe that a test case conducted within the DMFT calculation for the Hubbard model using the tensor trains impurity solver [citation needed] could further highlight this issue.

\subsection{Heating and Cooling}\label{subsec:HeatCool}

The possibility of obtaining a starting point for fixed-point iteration by extrapolating from a nearby temperature point allows to smoothly change temperature in subsequent simulation, in analogy to `heating' and `cooling' measurements in experiments. This capability is especially important at first-order phase transitions, where  multiple stable fixed points, corresponding to the different phases,  exist.

Such a heating and cooling process is shown in Fig.~\ref{fig:hysteresis}. Shown is a coexistence region and hysteresis curve at the example of the single-site DMFT on a Bethe lattice (the same calculation setup as the one mentioned in Sec.~\ref{subsec:convergence}). We plot the double occupancy, which is directly related to the potential energy, as a function of inverse temperature $\beta$. With starting points extrapolated from higher temperature converged results (red open circles), we find a transition from an insulating phase with small double-occupancy to a metallic phase with large double-occupancy around $\beta t= 21.56$. Conversely, with starting points extrapolated from lower temperature converged results (blue open circles), we find a transition from a metallic phase to an insulating phase around $\beta t= 20.31$. Between those temperatures, both metal and insulator are stable solutions of the fixed-point equations, indicating phase coexistence  \cite{georges1992,jarrell1992,rozenberg1992,zhang1993,rozenberg1999,bulla1999,Bluemer03,Erpenbeck2023}. %A precise analysis of the stability of this phase and its spinodal lines would require a thermodynamic analysis (rather than one based on the convergence of an iterative fixed point scheme) that is beyond the scope of this paper. \textcolor{magenta}{Should we put some reference here?}

\begin{figure}[H]
    \centering
    \includegraphics[width=1\linewidth]{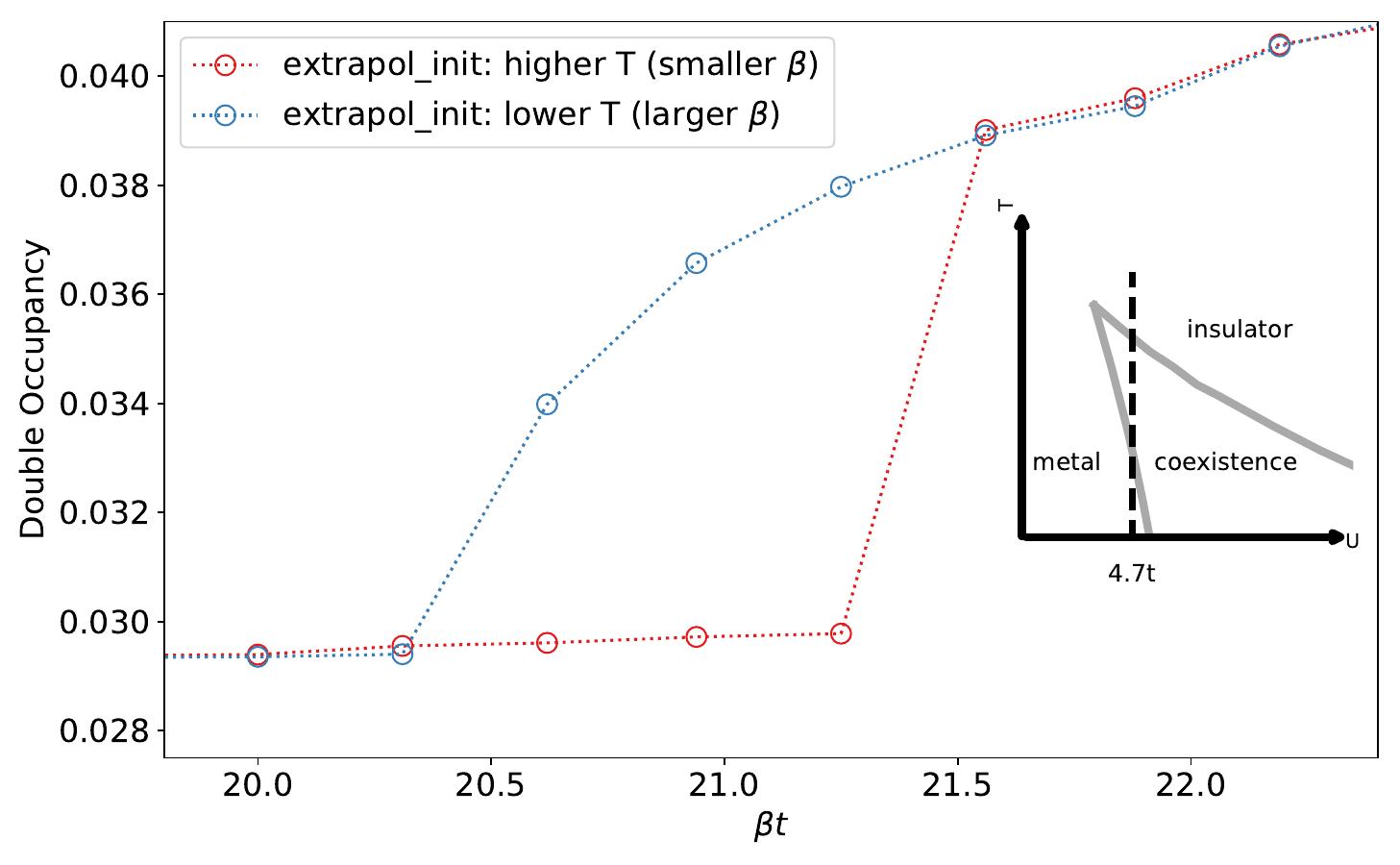}
    \caption{DMFT analysis of a half-filled Bethe lattice with $U=4.7t$ and a bandwidth of $4t$, displaying the hysteresis near $\beta t=21$. Points on the red curve utilize extrapolation from the nearest higher temperature converged results as the starting points for the calculation, while points on the blue curve employ extrapolation from the nearest lower temperature converged results as the starting points for the calculation. Inset: phase diagram adapted from Ref.~\cite{Bluemer03}.}
    \label{fig:hysteresis}
\end{figure}
\section{Conclusions}\label{sec:conclusions}
The Carath\'{e}odory temperature extrapolation technique, introduced in this paper, provides an effective way to accelerate the convergence of self-consistent many-body calculations for both model Hamiltonians and realistic systems. 
It allows for smooth temperature variation in simulations, making it suitable for studying heating and cooling processes in many-body systems.
 The starting points provided by this method are generally superior, especially in systems with phase transitions and convergence issues. 
 The Carath\'{e}odory technique therefore offers a versatile and efficient approach for studying temperature-dependent properties in self-consistent many-body calculations that should be adapted in any self-consistent finite-temperature many-body simulation. 

We note that `convergence acceleration' techniques such as direct inversion in the iterative subspace (DIIS)~\cite{pulay1980,pulay1982, pokhilko2022}  and Anderson acceleration~\cite{bacskay1981,walker2011,evans2020,pollock2021} are  complementary to the Carath\'{e}odory temperature extrapolation method employed in this study. Convergence acceleration techniques work by extrapolating from a set of initial iterations. The combination of such techniques with the better starting points provided by Carath\'{e}odory temperature extrapolation is therefore straightforward.

\begin{acknowledgments}
We thank André Erpenbeck for helpful discussions.
This material is based upon work supported by the National Science Foundation under Grant No. NSF DMR 2001465. This research used resources of the National Energy Research Scientific Computing Center (NERSC), a U.S. Department of Energy Office of Science User Facility located at Lawrence Berkeley National Laboratory, operated under Contract No. DE-AC02-05CH11231 using NERSC award BES-ERCAP0023196.
\end{acknowledgments}

\bibliographystyle{apsrev4-2}
\bibliography{ref}

\end{document}